\documentclass[twocolumn,showpacs,preprintnumbers,amsmath,amssymb]{revtex4}

\usepackage{graphicx}
\usepackage{dcolumn}
\usepackage{bm}
\newcommand{\be}{\begin{equation}}
\newcommand{\ee}{\end{equation}}
\newcommand{\bea}{\begin{eqnarray}}
\newcommand{\eea}{\end{eqnarray}}
\newcommand{\ba}{\begin{array}}
\newcommand{\ea}{\end{array}}
\newcommand{\bi}{\begin{itemize}}
\newcommand{\ei}{\end{itemize}}
\newcommand{\lan}{\langle}
\newcommand{\ran}{\rangle}

\begin{document}

\title{The nature and line shapes of charmonium in the $e^+e^- \to D\bar{D}$ reactions}

\author{Xu Cao$^{1,2,3}${\footnote{Electronic address: Xu.Cao@theo.physik.uni-giessen.de}} }
\author{H. Lenske$^{2,4}${\footnote{Electronic address: Horst.Lenske@theo.physik.uni-giessen.de}}}

\affiliation{$^1$Institute of Modern Physics, Chinese Academy of Sciences, Lanzhou 730000, China\\
$^2$Institut f\"{u}r Theoretische Physik, Universit\"{a}t Giessen, D-35392 Giessen, Germany\\
$^3$State Key Laboratory of Theoretical Physics, Institute of Theoretical Physics, Chinese
Academy of Sciences, Beijing 100190, China\\
$^4$GSI Darmstadt, D-64291 Darmstadt, Germany}
\date{\today}

\begin{abstract}

  We explore the nature of sharp resonances with asymmetric line shape observed in cross section data, a general physical phenomenon produced by the interference of continuum background and resonances. A Fano scheme and the coupled-channel T-matrix approach are employed to this aim and their close relationship is present. As a typical example, we point out that the $\psi(3770)$ state observed in the $e^+e^-$ reactions with an anomalous line shape can be explained naturally as a resonance embedded in the $D\bar{D}$ continuum. From a coupled-channels analysis the background of $\psi(3770)$ resonance is found to originate from a pole at $\sqrt{s}=3716.0 \pm$ 30.0~MeV. As a by-product, the broad structure $X(3900)/G(3900)$ seen in the Belle data, is found to be the tail of the $\psi(3770)$ state, distorted by the opening of the $D^*\bar{D} + c.c$ channel and the onset of the $\psi(4040)$ spectral distribution, thus making the assignment as a genuine charmonium state unlikely.

\end{abstract}
\maketitle{}

\section{Introduction} \label{introduction}

The proper treatment of near-threshold effects is a central issue of theoretical studies of charmonium spectroscopy. In fact, most of the experimental candidates for exotic $X,Y,Z$ states lie close to open-charm thresholds. A well known and intensively studied example is $X(3872)$, positioned within 1~MeV of the nominal $D^*\bar{D}$ threshold. The recently discovered structure $Z^+_c(3900)$ was observed first in the $J/\psi \pi^+$ invariant mass spectrum by the BESIII collaboration~\cite{BES:2013} and soon after confirmed by Belle~\cite{Belle:2013} and CLEO~\cite{CLEOc:2013}, respectively. Its mass is also close to $D^*\bar{D}$ threshold and the observed decay channel requires a $c\bar{c}d\bar{u}$ quark flavor structure component. Meanwhile, the BESIII collaboration has observed other charged charmonium-like states, namely $Z^+_c(4020)$~\cite{BES:1896} and $Z^+_c(4025)$~\cite{BES:2760}, waiting for confirmation by other experiments. Theses states contain the same flavor structure as $Z^+_c(3900)$ and are located near the $D^*\bar{D}^*$ threshold.

The internal structure and dynamical properties of a discrete quasi-bound state coupled to a continuum of unbound states are encoded in its line shape. Only in simple potential problems the line shapes of spectral distribution come close to the widely used Lorentz- or Breit-Wigner shapes, respectively. Under realistic conditions the shape of resonance is distorted by the interaction between the discrete and continuum components of the spectra. This kind of interference among states of various configurations is ubiquitous in quantum physics and leads to a plethora of interesting phenomena in nuclear, atomic, condensed matter and quantum optical physics. One of the recent exciting progress in this area is to control the atomic line shapes by the manipulation of the intense laser with different frequencies~\cite{Otto2013}. Obviously, this kind of direct manipulation is not within reach of hadron physics. The basically different situation in hadronic systems is their much shorter lifetime because of their strong interaction with neighbouring states. Hence, the line shapes of hadronic states are often found to be convoluted by overlapping contributions and, additionally, are influenced by coupled-channel effect, as e.g. in baryonic resonances~\cite{caoKSigma,JuliaDiaz:2006is,Matsuyama:2006rp}.

On the theoretical side, a profound approach accounting properly for interactions between discrete states and continuum was formulated a long time ago in a pioneering work by Fano~\cite{FanoPR1961}. While the original formulation was in the context of atomic physics, describing there the frequently observed phenomenon of the so-called auto-ionizing states, it was soon recognized that the Fano-mechanism is a phenomenon of compelling generality, appearing in any quantum system at any physical scale when interactions between discrete states and continuum occur. A more recent application to a nuclear physics problem is found in Ref.~\cite{Orrigo:2005}. With this paper, we intend to shed light on the special features encountered close to thresholds in the charmonium and bottonium region, closely following the idea promoted in our previous work~\cite{CaoFano2014}. Charmonium and bottonium states are of particular interest because of the appearance of sharp resonances with a width smaller by orders of magnitudes than the centroid mass, which is a defining feature of Fano resonances. In atomic and nuclear systems a closed channel is a transient phenomenon in the sense that at high enough energies it will turn into an open channel. The unique and new feature of charmonium and hadron physics in general is confinement, giving rise to the new phenomenon of absolutely closed channels, namely the hadronic states. Only to the expense of creating additional light ($q\bar q$) quark-antiquark pairs out of the vacuum the $Q\bar Q$ states can couple to the open $D\bar D$ channels. Interactions within the $Q\bar Q \leftrightarrow D\bar D$ system, or likewise the $(c\bar c) \leftrightarrow (c\bar q),(q\bar c)$ system, are leading to configuration mixing thus assigning charmonium and bottonium resonances an intrinsically complex structure. In production reactions we have to take into account channel coupling not only in the production amplitude but also configuration mixing in the produced charmonium states. One purpose of this paper is to point out the connection between the Fano-approach and the widely used coupled-channel methods, introduced already very early by the pioneering work of Eichten et al.~\cite{Eichten1980}. Such investigations have become of renewed interest \cite{klshnkv2005,Barnes2008} in connection with the newly found charmonium-like states.  The self-energies, induced by the coupling to the channels, are known to be essential for understanding the systematics of the charmonium spectrum~\cite{Barnes2008}. As the first charmonium state above the $D\bar{D}$ threshold, the $\psi(3770)$ state is ideal to investigate how this open charm coupling affects the spectral properties. In fact, its line shape observed in $e^+e^- \to D\bar{D}$~\cite{BESDDbar,BelleDDbar,CLEODDbar} and $e^+e^- \to hadrons$ \cite{BEShadrons,KEDRhadrons} cross sections, respectively, is anomalous, indicating coupled-channel effects. In previous studies \cite{HBLi2010,Zhang2010,YRLiu2010,Achasov2012,Chen2013} the major source of interactions was attributed to the $\psi(3686)$ state.

In section \ref{sec:FanoQuark} we review briefly the Fano-approach with special emphasis on applications to charmonium. The problem is considered from a general point of view in section \ref{sec:TheoryCC}. A coupled channels formalism for charmonium production is presented from which the Fano-description and the conventional T-matrix models are derived as limiting cases. For comparison a T-matrix coupled channels analysis of the $\psi(3770)$ data is performed in section \ref{sec:CharmCC} and results are being discussed. The paper closes by a summary and an outlook in section \ref{sec:Summary}.

\section{Fano-Coupling in Quarkonium}\label{sec:FanoQuark}

\begin{figure}
\begin{center}
{\includegraphics*[width=4.cm]{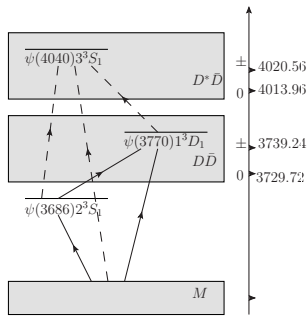}}
\caption{
Schematic representation of the transition from the non-resonant continuum built on the ground state $M$ to 1$^{--}$ charmonium states with two nearby continua $D^{(*)}\bar{D}$. The units of the charged ($\pm$) and neutral (0) thresholds are MeV.
\label{fig:FanoPic}}
\end{center}
\end{figure}

In this section we present a simple model for the treatment of configuration mixing interactions and their consequences for the physically observed charmonium spectrum. The scheme of states and continua underlying our considerations is depicted in Fig. \ref{fig:FanoPic}. We assume pre-diagonalization in the confined $c\bar c$ and of the open charm $c\bar q$ and $q\bar c$ sub-spaces, where the latter correspond to $D\bar D$ channels. Thus, a hybrid picture is used, considering on the one hand quarkonic $c\bar c$ dynamics and on the other hand hadronic $D\bar D$ dynamics. With respect to $c\bar c$ motion the confined states are closed channels at all energies while sub-threshold hadronic $D\bar D$ channels eventually change to open channels. Above the $D\bar D$ threshold, the by itself discrete spectrum of $c\bar c$ QCD-levels obtain a width by their coupling to the $D\bar D$ continua. Of course, this \emph{ansatz} is easily applied to any set of states, but here we consider specifically $\psi(3770)$ close to the $D\bar D$ threshold. For the sake of clarifying the basic mechanisms forming line shapes, in this section we neglect $D\bar D$ sub-threshold channels.

Suppose that we know the spectrum of bare $c\bar c$ states, their wave functions $\phi_c$ and mass $m_c$.  The same is assumed for the hadronic $D\bar D$ scattering states with relative motion wave functions $\phi_{d}(\omega)$ at energy $\omega=\sqrt{s}$. As in the original work of Fano \cite{FanoPR1961}, the eigenstates of the interacting system are expanded into the subsets of state vectors, as introduced in our previous work \cite{CaoFano2014}. For clarity, we consider the simplest case given by a bare state $c$ and a single continuum of the state $d$. The state vector is
\be \label{eq:FanoWave}
\Psi_\omega =z_c(\omega)\phi_c(\omega)+\int{d\omega' z_{d}(\omega')\phi_{d}(\omega')} \quad .
\ee
which separates intermediate propagators into pole and principal values contributions, considering that the coupling to closed channels lead to dispersive, but not absorptive self-energies, as shown in the original Fano paper \cite{FanoPR1961}. Interactions modify the wave function of resonance to be dressed by a cloud of virtual $D\bar D$ continuum
\be
\chi_{c}(\omega)=\phi_c(\omega) +P\int{d\omega'\frac{V_{cd}}{\omega-\omega'}\phi_{d}(\omega')}
\ee
and the correlated state vector is obtained as
\be
\Psi_\omega=x_c(\omega)\chi_{c}(\omega)+x_d(\omega)\phi_{d}(\omega)
\ee
Here, $V_{cd}(\omega)$ denotes the matrix element of the configuration mixing interaction. The interaction may lower one or a few eigenstates below the particle emission threshold, and one may speculate whether $\psi(3686)$ is of such a nature, see e.g. Ref.~\cite{klshnkv2005}. The amplitudes $x_{c,d}$ are obtained for the solution of a set of coupled equations and by the proper normalization of the state vector \cite{FanoPR1961,Orrigo:2005,CaoFano2014}. Their detailed forms are of no special interest here. A more important message of Eq.~(\ref{eq:FanoWave}) is that the observed charmonium states like $\psi(3770)$ have to be considered as varying mixtures of $c\bar c$ and $D\bar D$ configurations. While the bare $c\bar c$ states by itself lives indefinitely long, the configuration interactions $V_{cd}$ induce a spectral distribution of a width $\Gamma_c(\omega) = 2\pi |V_{cd}(\omega)|^2$ and a related energy dependent mass shift $\Delta m_c(\omega)$. The coupling of bare state to continuum induces an additional configuration mixing phase shift, derived in the present context as
\be\label{eq:deltaMIX}
\tan{\delta_{cd}}(\omega)=\frac{x_c(\omega)}{x_d(\omega)}=\frac{m_c\Gamma_c(\omega)}{m^2_c-\omega^2}
\ee
thus recovering the well-known Breit-Wigner relation for a resonance phase shift, obtained here as a result of configuration mixing. Above, $m_c=m^0_c+\Delta m_c$ includes the mass shift. However, because of the extremely small width, $\Gamma_c \ll m_c$, both $m_c$ and $\Gamma_c$ can be taken as constant in the resonance region. The phase shift $\delta_{cd}$ varies rapidly with energy on a scale set by $V_{cd}\sim \sqrt{\Gamma_c}$. The configuration mixing phase shift $\delta_{cd}$ has to be added to the bare $D\bar D$ channel phase shift varying on a much larger energy scale, given by $t$- and $u$-channel interactions from the exchange of light mesons. Hence, in (hypothetical) $D\bar D$ scattering one would observe a sharp resonance around $\omega \sim m_c$, superimposed on and interfering with a slowly varying background.

In order to work out the role played by $c\bar c$ states in the resonances observed in $e^+e^-$-annihilation reactions we need to combine properly the reaction model on the one side and the configuration model on the other side. Starting from an initial reaction channel $|\tau \rangle$ at total energy $\omega$, let be $M_\tau$ the transition operator for the production of the state $\Psi_\omega$. In the following formulae we omit partial wave indices because we are studying the production and decay of $1^{--}$ charmonium vector states which couple to $D\bar D$ $P$-waves. Obviously, the formalism is easily extended to any other partial waves. The charmonium production amplitude out of the incident channel $|\tau \rangle$ is described by the matrix element of the corresponding operator $\mathcal{M}_\tau$
\be \label{eq:ProdAmp}
\langle \Psi_\omega|\mathcal{M}_\tau|\tau\rangle=x_c(\omega)
\langle \chi_c|\mathcal{M}_\tau|\tau\rangle+x_s(\omega)\langle \phi_d|\mathcal{M}_\tau|\tau\rangle
\ee

The reaction amplitude is given by a production form factor which we express as \cite{FanoPR1961,CaoFano2014}
\be \label{eq:FanoFactor}
|F^D_{\tau,d}|^2=|\lan \phi_d|\mathcal{M}_\tau|\tau\ran|^2\frac{|q_{c\tau}-\varepsilon|^2}{1+\varepsilon^2} \quad ,
\ee
where $\varepsilon=cot\delta_{cd} = (-\omega^2+m^2_{c})/m_{c}\Gamma_{c}$ is due to configuration mixing. Thus, we have obtained an expression going beyond the standard Breit-Wigner parameterizations of line shapes. The new ingredient is the quantity
\be\label{eq:FanoPar}
q_{c\tau}(\omega)= \frac{\lan \chi_c|\mathcal{M}_\tau|\tau\ran}{\lan \phi_d|\mathcal{M}_\tau|\tau\ran}
\ee
which plays a central role in our approach in controlling the line shape. In general, $q_{c\tau}$ is a complex-valued function of nergy and momentum where $|q_{c\tau}|^2$ is the a measure of the population probability of the (dressed) $c\bar c$ state relative to the population probability of the purely hadronic $D\bar D$ channel. To a very good approximation we are allowed to use $q=q_{c\tau}(m_{c})=const.$ for narrow resonances. Eq.~(\ref{eq:FanoFactor}) shows that $q_{c\tau}$ is controlling the line shape of the spectral distribution: a dip, eventually down to zero, will appear at an energy $\omega_0$ where $Re(q_{c\tau})=\varepsilon(\omega_0)$. For $q_{c\tau}=0$ an inverted resonance line shape with a minimum at $\omega=m_c$ will occur. The widely used Breit-Wigner profile is recovered if $|q_{c\tau}| \gg \varepsilon$ over the whole resonance region, leading to:
\be
F_{c}(s) = \frac{A_{c}}{s-m^2_{c}+i m_{c}\Gamma_{c}} \quad .
\label{eq:BWpsi3770}
\ee
The latter two cases correspond to the limiting scenarios of the reaction, namely for $q_{c\tau}=0$ exclusive annihilation into the hadronic $D\bar D$ channel and, as the other extreme, exclusive annihilation into the $c\bar c$ channel for $|q_{c\tau}|\to \infty$. The latter case describes, by the way, sub-threshold charmonium production for which the form factor, Eq.~(\ref{eq:FanoFactor}) reduces to the $c\bar c$ amplitude. Thus, besides fixing the line shape, $q_{c\tau}$ provides information on the reaction mechanism. As such, it depends naturally on the type of reaction and we have to expect different line shapes when populating \emph{the same final states} in different reactions. Thus, spectral distribution of different shapes have to be expected in charmonium production in leptonic $\ell\bar \ell$  and hadronic $h\bar h$ annihilation reactions. If many open channels are contributing, the interference minimum will be superimposed on a finite background. The structural properties of charmonium are imprinted in $\varepsilon$, given by the configuration mixing phase shift $\delta_{cd}$. Hence, $\varepsilon$ contains the full spectral information on the mass and the width of the resonance.

In the specific case of $c = \psi'(3770)$ the width is given by the $P$-wave relation \cite{Zhang2010}
\be\label{eq:Width3770}
\Gamma_c=\Gamma_{\psi'}(s) =\frac{g^2_{\psi'D\bar{D}}}{6\pi s} \left(p^3_0(s) + p^3_\pm(s)\right )
\ee
where the c.m. momenta are $p_{0,\pm}(s) = \sqrt{s/4-m^2_{D^{0,\pm}}}$ for neutral and charged $D\bar{D}$ channels. Finally, the yet missing population probability of the hadronic $D\bar D$ component in Eq.~(\ref{eq:FanoFactor}) is defined by the \emph{ansatz}
\be
|\lan \phi_d|\mathcal{M}_\tau|\tau\ran|^2 = | A_{\psi'}F_{d}|^2
\ee
The energy dependence is described by the form factor
\be\label{eq:FormDD}
F_{d}(s) = \frac{1}{s-m^2_{d}+im_{d}\Delta_{d}} \quad ,
\ee
mimicking the effect of the background contributions, which mainly come from $\psi(3686)$ state as shown by several studies \cite{HBLi2010,Zhang2010,YRLiu2010,Achasov2012,Chen2013}. The parameters extracted from data are consistent with our expectation and seem to confirm our choice. The Breit-Wigner form has been chosen for convenience, but other functional forms may be used as well. Hence, the quantities $m_d$ and $\Delta_d$, listed in Tab.\ref{tab:fitparaFano}, are purely phenomenological quantities of mainly numerical character. Different from the resonance parameters discussed below, their values might vary with the energy interval considered in the fit. The magnitude of the $e^+e^- \to D\bar D$ production amplitude is fixed by $A_{\psi'} = m^2_{\psi'} g_{\psi'D\bar{D}}/g_{\psi'\gamma}$, determined essentially by the $g_{\psi'D\bar{D}}$, dimensionless coupling constant of $\psi(3770)$ to $D\bar{D}$.

For the present investigations, we determine the photo-vector coupling constant $g_{\psi'\gamma}$ phenomenologically. The electronic width of vector charmonium states is given by $\Gamma_{\psi' e^+e^-} = {4\pi \alpha^2 m_{\psi'} }/{3 g^2_{\psi'\gamma}}$, see e.g. \cite{Novikov:1978} where $\alpha \simeq 1/137$ denotes the electromagnetic fine structure constant. With  $\Gamma_{\psi' e^+e^-} = 0.265$~keV from the recent compilation of the Particle Data Group~\cite{pdg2010} the photo-vector coupling constant  at $m_{\psi'}=3770$~MeV was determined to be $g_{\psi'\gamma}=56.35$.

We define the continuum open charm production cross section, including the appropriate two-body phase space factor,
\be
\sigma_{D\bar D}(s) = \frac{8\pi \alpha^2 p^3_{d}}{3s^{5/2}}|\lan \phi_d|\mathcal{M}_\tau|\tau\ran|^2 \quad .
\ee
with $p_d=p_{0,\pm}$, depending on the channel. We obtain the full $e^+e^- \to D\bar D$ cross section as
\be\label{eq:XSpsi3770}
\sigma_d(s) = \frac{8\pi \alpha^2 p^3_{d}}{3s^{5/2}} |F^D_{\tau,d}|^2 =\sigma_{D\bar D}(s)\frac{|q-\varepsilon|^2}{1+\varepsilon^2} \quad .
\ee
Hence, the charmonium production cross section is separated into the annihilation cross section populating the hadronic $D\bar D$ component and a form factor containing the population and spectroscopy of the confined $c\bar c$ component of the full charmonium state vector.

\begin{figure}
\begin{center}
{\includegraphics*[width=9.cm]{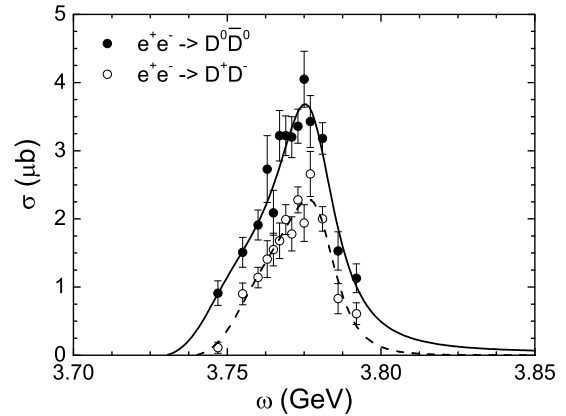}}
\caption{
Total cross section of $e^+e^- \to D^0\bar{D}^0$ and $e^+e^- \to D^+D^-$ charmonium production. The results obtained with Eq.~(\ref{eq:XSpsi3770}) for the $D^0\bar D^0$ (solid line) and the $D^+D^-$ channel (dashed line), respectively, are compared to the data are from the BES collaboration~\protect\cite{BESDDbar}.
\label{fig:FanoFit}}
\end{center}
\end{figure}

Applying the approach to the BESIII data~\cite{BESDDbar}, the spectral distributions in the $D^0\bar D^0$ and the $D^+D^-$ production cross sections are well described, as seen in Fig.~\ref{fig:FanoFit}. In order to illustrate applications to data analyses, the mass, width and line shape parameters, respectively, have been varied freely in a $\chi^2$-~minimization process. The resulting parameter sets are shown in Tab.~\ref{tab:fitparaFano}. The bare mass and width of $\psi(3770)$ in the neutral and charged channels are consistent with each other. The different behaviour of the two production cross sections results almost totally from the difference in phase space factors of the $D^0\bar{D}^0$ and $D^+D^-$ channels, namely the mass gap of neutral and charged $D$-meson. It should be mentioned that the dip at around 3.82 GeV could be reproduced with the above prescription if the Belle data at high energies are included in the fit, as we have shown previously \cite{CaoFano2014}. Finally, attempting a fit with a simple Breit-Wigner line shape (i.e. assuming $|q|\to \infty$) the description deteriorates as reflected by the increased values $\chi^2= 2.72$ for the $D^0\bar D^0$ channel and $\chi^2=3.27$ for the fit to the $D^+D^-$ data, respectively.

The slightly different results for $q$ obtained from the $D^0\bar D^0$ and the $D^+D^-$ data agree within the error bars but they may taken as an indication on the reaction-dependence of spectral line shapes observed in production reactions. Despite the remaining uncertainties due to the relatively large experimental errors the two $q$-values are indicating differences in the reaction mechanism, probably mainly caused by differences in the final states interactions. More precise data from future experiments, either at $e^-e^+$-facilities or from $p\bar p$ annihilation as planned at PANDA@FAIR are important for a more detailed analysis.

\begin{table}[t]
\begin{center}
\begin{tabular}{c|c|c}
$c=\psi'$      & $D^0\bar{D}^0$ & $D^+D^-$ \\
\hline
$m_{\psi'}$~(MeV) & $3782.1 \pm 1.6$ & $3784.0 \pm 2.0$ \\
$g_{\psi'D\bar{D}}$ & $11.8 \pm 0.9$ & $10.7 \pm 1.3$ \\
$q$ & $-2.1 \pm 0.3$ & $-1.6 \pm 0.3$ \\
$m_{d}$~(MeV) & $3743.0 \pm 5.4$ & $3753.3 \pm 3.9$ \\
$\Delta_{d}$~(MeV) & $34.1 \pm 5.2$ & $33.3 \pm 5.6$ \\
$\chi^2/d.o.f$ & $0.83$ & $0.90$ \\
\end{tabular}
\end{center}
\caption{Charmonium production in the $e^+e^- \to D^0\bar{D}^0$ and $e^+e^- \to D^+D^-$ annihilation reactions: Fano-parameters according to Eq.~(\protect\ref{eq:FanoFactor}) from the fit to the data shown in Fig.\protect\ref{fig:FanoFit}.
\label{tab:fitparaFano}}
\end{table}

\section{Coupled Channels Approach to Charmonium Production}\label{sec:TheoryCC}

The primary goal of this section is to work out on a formal level the connections between the Fano-formalism and standard coupled channels theory. While the latter has been considered extensively in the literature, see e.g. \cite{Aceti:2014ala,Hyodo:2011qc,Hyodo:2007jq,Hyodo:2013nka}, the special aspect introduced by the interference of open and closed channels are deserving a closer discussion. The major results are found in Sect. \ref{subsec:Relation} to which readers acquainted with formal coupled channels scattering theory may go directly. We present an approach which on equal footing treats the production of confined $c\bar c$ and $D\bar D$ states without distinguishing sub- and above-threshold production. The two types of configuration are, in principle, present in any charmonium states, irrespective of the mass, because of configuration mixing. The amount of mixing, however, will change with energy. While below threshold $D\bar D$ states contribute only as virtual admixtures to $c\bar c$ charmonium, it changes at higher energies to observable on-the-mass-shell components as detected in experiments.

In order to account for the multiple facets of charmonium, we coalesce the discrete $c\bar c$ confinement sector and the $ D\bar D$ continuum sector into a unified approach. The confined $c\bar c$ states and the open charm $D\bar D$ configurations are arranged in disjoint, orthogonal subspaces $C$ and $D$, respectively. The $C$ and $D$ sectors are assumed to be pre-diagonalized. Hence, $C$ contains bare $c\bar c$ states confined into a meson-like structure, and $D$ contains the $c\bar q$ and $q\bar c$ configurations, condensed in $D$ and $\bar D$ mesons. The $c\bar c \to D\bar D$ is proceeded by cutting the  $c\bar c$ string with the presence of the additional $q\bar q$ pair. This happens to the expense of creating new confinement bonds appearing as the $D$ and $\bar D$ mesons. The latter are allowed to interact by long range, van-der-Waals-type strong forces and eventually electromagnetic interactions as typical for low-energy hadron physics.

Throughout this paper, the $c\bar c$ states are closed channels with respect to the quark motion, constraint by confinement into a non-separable mesonic system. Of course, in the lab-system the $c\bar c$ entity moves as a whole and without coupling to decay channels, it will be a stable particle except for weak and electromagnetic interactions which we neglect here. The $D\bar D$ channels represent the quasi-free motion of the $c$ quark and the $\bar c$ anti-quark, separated because of their dressing by a light $q \bar q$ quark-antiquark pair, where $q\in\{u,d,s\}$. Obviously, this view is slightly different from the one underlying the commonly used in T-matrix approaches. They are emphasizing the $D\bar D$ content of charmonium. The resonances formed in $D\bar D$ scattering are described typically by phenomenologically introduced, but otherwise unresolved vertices as, for example, in \cite{Achasov2012,Hanhart2012,Zhang2010}. Here, we take the point of view that $c\bar c$ states are existing by their own right, as it is the case below the $D\bar D$ threshold. Above threshold, the $c\bar c$ states are coupled to the $D\bar D$ open charm sector and obtain a finite lifetime and spectral width. This picture is occasionally used in descriptions of charmonium spectroscopy inspired by the quark model, see e.g. \cite{klshnkv2005}.

\subsection{Theory of Charmonium Production Reactions} \label{subsec:charmth}

The configuration space is divided into the subspaces  $C$ and $D$. The confined $c\bar c$ states are contained in the space $C$ while the space $D$ is composed of the $(c\bar q)(q\bar c)$ configurations, condensed into $D\bar D$ mesons. The (multi-channel) propagators within the separate $C$ and $D$ sectors are denoted by $G^C$ and $G^D$, respectively. Together, they are forming the bare propagator in the combined space, $G_0=diag(G^C,G^D)$. Two spaces interact via the residual interaction $V$, represented by a matrix with only non-diagonal elements $V^{CD}$ and $V^{DC}$, respectively. The propagator $G=G_0+G_0VG$ of the fully interacting system is
given by a two-by-two matrix structure. With the M{\o}ller-vertex
\bea \label{eq:GammaA}
\tilde{\Gamma}^A = \Gamma^{A,out}=\left(1-\Sigma^A G^A\right)^{-1}=1+\Sigma^A G^{AA}
\eea
with $A\in\{C,D\}$ here and afterwards. Due to time-reversal invariance the incoming and outgoing vertices are related by vertex $\Gamma^{A,in,\dag} = \Gamma^{A,out}$. The diagonal coupled channels propagators are given by
\be
G^{CC}=\tilde{\Gamma}^{C}G^C
; \quad
G^{DD}=\tilde{\Gamma}^{D}G^D
\label{eq:GreenAA}
\ee
and the non-diagonal pieces of $G$ are related to the diagonal ones:
\be
G^{CD}=G^CV^{CD}G^{DD}; \quad G^{DC}=G^DV^{DC}G^{CC \label{eq:GreenCD}}
\ee
The self-energies
\be
\Sigma^C=V^{CD}G^DV^{DC}; \quad \Sigma^D=V^{DC}G^CV^{CD} \label{eq:SigmaAA}
\ee
describe the induced dispersive interactions. At energies above the $D\bar D$ threshold the self-energies $\Sigma^C$ are non-hermitian because they are induced by the open $D\bar D$ channels. Hence, the otherwise sharp, non-decaying $c\bar c$ states gain an energy dependent mass shift and a decay width, leading to a finite life time, provided their bare mass exceeds the lowest open charm threshold. Because of confinement the $c\bar c$ states do not carry asymptotic flux in the center-of-mass (c.m.) frame and therefore they contribute only an energy dependent dispersive correction to the elastic $D\bar D$ interactions.

For the further steps it is of advantage to split the self-energies in subspace $A$ into their diagonal parts $U^A$ and their configuration mixing parts $W^A$, defined by
\bea
U^A_{ij}&=&\Sigma^A_{ii}\delta_{ij} \label{eq:SigmaU}\\
W^A_{ij}&=&\Sigma^A_{ij}-U^A_{ij}   \label{eq:SigmaW} .
\eea
Similar decomposition has been used to explore the compositeness of states in conventional hadronic reactions and effective potential~\cite{Aceti:2014ala,Hyodo:2011qc,Hyodo:2007jq,Hyodo:2013nka}. Here we define the purely diagonal vertex
\be \label{eq:VertexAU}
\Gamma^A=\left(1-U^AG^A\right)^{-1} ,
\ee
and the vertex correlation matrix containing non-diagonal and diagonal interactions
\be \label{eq:VertexCO}
\Omega^A=\left(1-W^A G^A \Gamma^A \right)^{-1} .
\ee
We obtain the representation:
\be
\tilde{\Gamma}^A=\Gamma^A \left(1-W^AG^A\Gamma^A \right)^{-1}=\Gamma^A\Omega^A .
\ee
thus separating formally re-scattering effects leading back to the initial state from those connecting inelastically different states within each of the $C$ and $D$ subspaces. Since the propagators $G^A \Gamma^A$ include self-energies, they account for effects responsible for the finite life times and physical masses of the $C$ configurations.

Turning now our attention to the production reaction $|\tau\ran\to|\Psi_{\omega}\ran$ we define the reaction form factor $F_{\tau}=diag(F_{\tau}^C,F_{\tau}^D)=M_\tau+VGM_\tau$. Both $M_\tau$ and $F_{\tau}$ are having formally a two-component vector structure. The re-scattering matrix is
\be
1+VG=
\begin{pmatrix}
  1+V^{CD}G^{DC}&V^{CD}G^{DD}\\
  V^{DC}G^{CC}&1+V^{DC}G^{CD}
\end{pmatrix}\\
\ee
Specifically, we consider $\tau\equiv e^-e^+$ and the charmonium state $\Psi_{\omega}$, composed of the $c\bar c$ and $D\bar D$ components. Different from other approaches~\cite{Aceti:2014ala}, we have used the Low-equation for expressing the re-scattering part in terms of the tree-level interaction $V$ and the dressed Green's function $G$ rather than the T-matrix and the bare propagators. We find
\bea
F_{\tau}^C &=& \Gamma^C \Omega^C M^C_{\tau} + V^{CD} G^D \Gamma^D \Omega^D M^D_{\tau} \quad \\
F_{\tau}^D &=& \Gamma^D \Omega^D M^D_{\tau} + V^{DC} G^C \Gamma^C \Omega^C M^C_{\tau} \quad
\eea
Inserting the complete set of eigenstates $|\alpha_c\ran$ and $|\alpha_d\ran$ of dressed propagators $G^{A} \Gamma^A$, respectively, at appropriate places, we project $F_\tau$ at given c.m.-energy $\omega=\sqrt{s}$ from the left onto the channel state vector  $\lan\Phi_\omega|=(\lan\phi_c|,\lan \phi_d|)$, composed of eigenstates of the bare propagators $G^{C,D}$ of the uncoupled system. Those states are making up the asymptotically separated system. From the right, we project onto the initial channel $|\tau\ran$. As a result, we find the reaction amplitude $f_{\omega\tau}=\tilde{f}^C_{\omega\tau}+\tilde{f}^D_{\omega\tau}$. Denoting the overlap matrix elements of the bare and the coupled eigenstates by the configuration amplitudes $x_{a\alpha_b}(\omega)=\lan\phi_a|\alpha_{b}\ran$ we find
\bea
\tilde{f}^D_{\omega\tau}&=&\sum_{\alpha_c}x_{c\alpha_c}[\sum_{\beta_c}{\Gamma^C_{\alpha_c}\Omega^C_{\alpha_c\beta_c}M^C_{\beta_c\tau}} \nonumber \\
&+&\sum_{\alpha_d,\beta_d}{V^{CD}_{\alpha_c\alpha_d}G^D_{U,\alpha_d}\Omega^D_{\alpha_d\beta_d}M^D_{\beta_d\tau}}]\quad ,\\
\tilde{f}^D_{\omega\tau}&=&\sum_{\alpha_d}x_{d\alpha_d}[\sum_{\beta_d}{\Gamma^D_{\alpha_d}\Omega^D_{\alpha_d\beta_d} M^D_{\beta_d\tau}} \nonumber \\
&+&\sum_{\alpha_c,\beta_c}{V^{DC}_{\alpha_d\alpha_c}G^C_{U,\alpha_c}\Omega^C_{\alpha_c\beta_c}M^C_{\beta_c\tau}}]
\quad \label{eq:fftoFF},
\eea
describing the production of charmonium states which are eigenstates of the interacting $c\bar c \leftrightarrow D\bar D$ system.

\subsection{Relation to the Fano- and  other Approaches}\label{subsec:Relation}

\begin{figure}
\begin{center}
{\includegraphics*[width=8.cm]{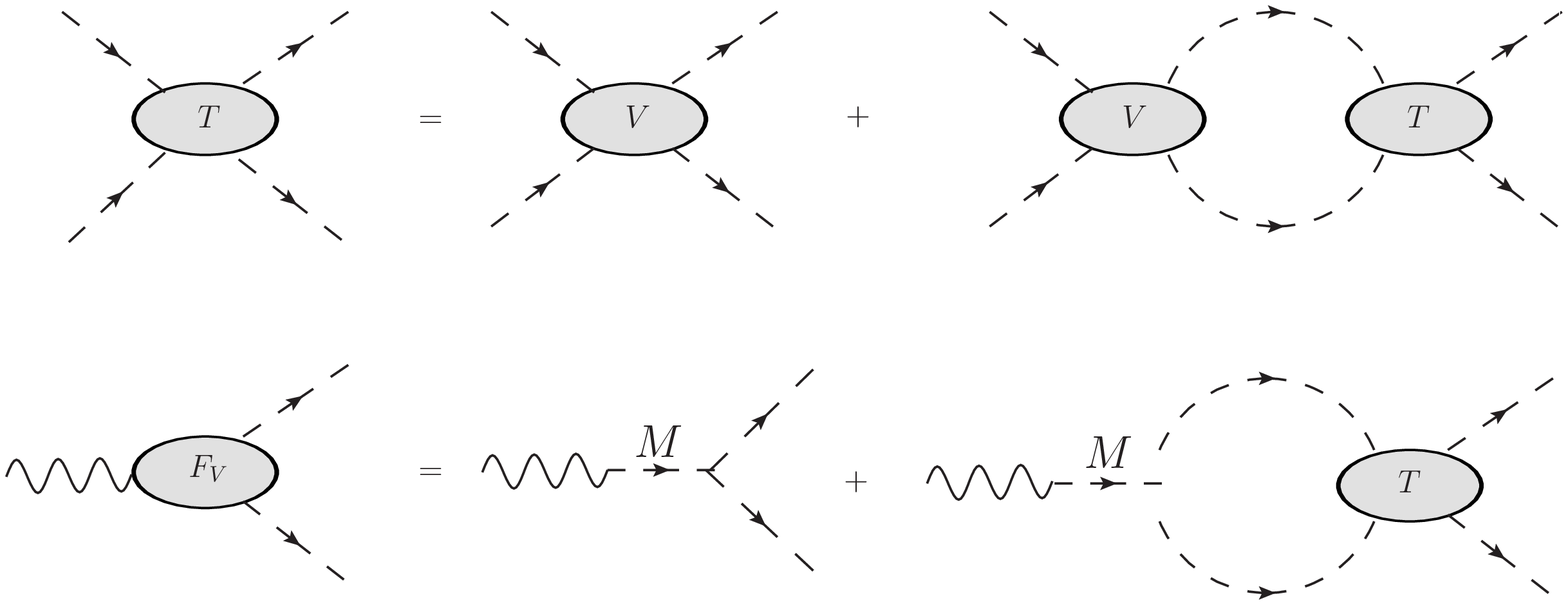}}
\caption{
Graphical representation for the $D^{(*)}\bar{D}$ scattering (top) and electromagnetic form factor of $\psi(3770)$ (bottom).\label{fig:feydiagr}
}
\end{center}
\end{figure}

The probabilities to find the total amount of $C$-type and the $D$-type configurations, respectively, are
\be
|z^C_\omega|^2=\sum_{\alpha_c}{|x_{c\alpha_c}|^2}\quad ;\quad |z^D_\omega|^2=\sum_{\alpha_d}{|x_{d\alpha_d}|^2}
\ee
and by means of the normalized amplitudes
\be
f^C_{\omega\tau}=\frac{1}{z^C_\omega}\tilde{f}^C_{\omega\tau}\quad ;\quad
f^D_{\omega\tau}=\frac{1}{z^D_\omega}\tilde{f}^D_{\omega\tau}
\ee
the complete reaction form factor is obtained as
\be
f_{\omega\tau}=z^C_\omega f^C_{\omega\tau}+z^D_\omega f^D_{\omega\tau}
\ee
Of primary interest is the detectable $D\bar D$ component which is contained in $f^D_{\omega\tau}$. Therefore, we write in analogy to the Fano-formula
\be
f_{\omega\tau}=f^D_{\omega\tau}z^C_\omega\left(q^{CD}+\varepsilon^{CD} \right)
\ee
where
\be
q^{CD}=\frac{f^C_{\omega\tau}}{f^D_{\omega\tau}}\quad ; \quad \varepsilon^{CD}=\frac{z^D_\omega}{z^C_\omega}
\ee
and taking into account $|z^C_\omega|^2+|z^D_\omega|^2=1$ we find
\be
f_{\omega\tau}=f^D_{\omega\tau}e^{i\chi_{C}}\frac{q^{CD}+\varepsilon^{CD}}{\sqrt{1+|\varepsilon^{CD}|^2}}
\ee
where $\chi_{C}$ denotes the phase of $z^C_\omega$. Thus, we have derived on the general grounds of couple channels theory a Fano-type formula, covering the general case of an arbitrary number of discrete states and continua and establishing the relation to coupled channels scattering theory. Moreover, the results confirm the Fano-conjecture that interference structures in spectral distributions of long-living states give access to a quantitative determination of configuration mixing. The above formulas show that the comparison of characteristic line shape features observed in production reactions with different initial channels corresponds to scan configuration mixing effects in the produced states. In the case of a single state and one continuum background and minor rearrangement of terms, we recover the Fano-result Eq.~(\ref{eq:FanoFactor}) discussed in the previous section, noting that under those conditions the vertex correlation matrices are reducing to be identity, $\Omega^A\to I$.

A different scheme is used in $T$-matrix approaches, specifically focussing on the $D\bar D$ exit channels, but also easily extended to include higher open charm channels. Theoretically, such prescriptions mean to project the full amplitude onto the channel state vector $\Phi_d\equiv (0,\phi_d)^T$ containing a selected, specific single $D\bar D$ configuration as observed in the detector. However, from a strict point of view such a restricted \emph{ansatz} is not fully justified because it is biased by the assumption that above the $D\bar D$ threshold charmonium states are completely given by $D\bar D$ configurations. Putting behind these objections for the moment, we find
\be
F^D_{\tau} = {f}^D_{\omega\tau}
\label{eq:TDD}
\ee
The physical meaning of above equation is seen clearly by considering the structure of Eq.~(\ref{eq:fftoFF}). Using Eq.~(\ref{eq:GammaA}) we have
\bea
F^D_{\tau}&=&\tilde{\Gamma}^DM^D_\tau+V^{DC}G^C\tilde{\Gamma}^CM^C_\tau \nonumber \\
&=&\left(1+\Sigma^DG^{DD}\right)M^D_\tau \nonumber  \\
&+& V^{DC}G^C\left(1+\Sigma^CG^{CC}\right)M^C_\tau \label{eq:OP_FDD}
\eea
The first term describes the direct $D\bar D$ production and re-scattering of the mesons on the induced self-energies. The second term actually plays the role of a vertex correction, as depicted in Fig.~\ref{fig:feydiagr}. We introduce the effective $D\bar D$ vertex
\be\label{eq:effVertex}
\tilde{M}^D_\tau\equiv M^D_\tau+V^{DC}G^CM^C_\tau \quad .
\ee
With repeated use of Eq.~(\ref{eq:SigmaAA}), expressing $G^{C,D}$ in terms of $G^{CC,DD}$ by means of Eq.~(\ref{eq:GreenAA}), and after a suitable rearrangement of terms, the last term at the very right end of Eq.~(\ref{eq:OP_FDD}) is transformed into
\be
V^{DC}G^C\Sigma^CG^{CC}= \Sigma^DG^{DD}V^{DC}G^C+\mathcal{O}(V^4)
\ee
Hence, up to $\mathcal{O}(V^4)$, Eq.~(\ref{eq:OP_FDD}) is obtained in the rearranged form
\be \label{eq:FDD}
F^D_{\tau}\simeq \tilde{M}^D_\tau+\Sigma^DG^{DD}\tilde{M}^D_\tau=\tilde{M}^D_\tau+T^{D\bar D}G^{D}\tilde{M}^D_\tau
\ee
as displayed in the lower panel of Fig.~\ref{fig:feydiagr}. Here we have introduced the $D\bar D$ T-matrix
\be
T^{D\bar D}=\Sigma^D+\Sigma^DG^DT^{D\bar D}
\ee
Obviously, Eq.~(\ref{eq:FDD}) is equivalent to
\be \label{eq:FormDD2}
F^D_{\tau}=\Gamma^D\Omega^D\tilde{M}^D_\tau
\ee
with $\Omega^D$ in Eq.~(\ref{eq:VertexCO}). In the next section, $D\bar D$ production will be investigated by using this representation.

\section{Coupled Channels T-Matrix Analysis of Charmonium Production}\label{sec:CharmCC}

In Sec.~\ref{sec:FanoQuark}, we have given a Fano picture of the $\psi(3770)$ state. For a more extended description of the physical situation in Fig.~\ref{fig:FanoPic}, covering a larger energy interval, a treatment of the multi-channel case, e.g. $D\bar{D}$ and $D^{*}\bar{D}+c.c.$ is necessary. The Fano-approach, in fact, is flexible enough to achieve that goal, finally coming down to the solution of the coupled-channel problem for any number of states and channels \cite{FanoPR1961,Orrigo:2005}. It also has the advantage to explore the confined heavy quark-antiquark states and asymptotically scattering states in a consistent way~\cite{CaoFano2014}. However, here we use directly more conventional coupled-channel theory demonstrated in Sec.~\ref{sec:TheoryCC}, which is easily applied to multi-channel cases, as e.g. in our recent work on strangeness production on the nucleon~\cite{caoKSigma}. The formalism is easily extended to non-$D\bar{D}$ channels which play a key role in understanding the nature of the $\psi(3770)$.

Quarkonia have been studied with coupled-channel methods by various authors, for recent works see e.g. \cite{Achasov2012,zhangPRL09,Wang:2011yh}. Being a $P$-wave state, the formation of $\psi(3770)$ in $D\bar D$ scattering resembles the formation of $\rho(770)$ as an isovector $\pi\pi$ $P$-wave resonance. The latter was studied recently by Hanhart \cite{Hanhart2012}, using a coupled channels approach to investigate contributions of other resonances to the form factor of $\rho(770)$. Here, we follow closely the scheme developed in the previous section. Our starting point is Eq.~(\ref{eq:FormDD2}). As proved in the previous section the set of equations to be solved is in the representation
\bea \label{eq:newFDD2}
F^D_{\tau}&=&\Gamma^D(1-W^DG^D\Gamma^D)^{-1}\tilde{M}^D_\tau
\eea
together with $\Omega^D$ in Eq.(~\ref{eq:VertexCO}) and $\tilde{M}^D_\tau$ in Eq.~(\ref{eq:effVertex}).
According to Eq.~(\ref{eq:SigmaW}) the self-energy $\Sigma^D$ is split into the diagonal part $U^D$ and the configuration mixing part $W^D$, respectively. The $W^D$ is purely non-diagonal in conventional decomposition, see e.g.~\cite{JuliaDiaz:2006is,Matsuyama:2006rp}. Following our alternatively practice of Ref.~\cite{caoKSigma}, but also used elsewhere, e.g. Ref.~\cite{Hanhart2012}, herein we consider $U^D$ as pure elastic part and $W^D$ as resonant part. Concretely, only $\psi(3770)$ is included into $U^D$ and $\Gamma^D$ as a $D\bar{D}$ elastic resonance and other charmonium states, e.g. $\psi(3686)$ and $\psi(4040)$ with weaker coupling to the $D\bar{D}$ channel are included in the $W^D$ as inelastic resonances. This amounts to consider $\psi(3770)$ as a $D\bar{D}$ elastic resonance in $s$-channel which is well justified because of the weak non-$D\bar{D}$ decay of $\psi(3770)$, resembling also in this aspect the $\rho(770)$ with its dominant decay into the $\pi \pi$ isovector $P$-wave channel. Correspondingly, the $W^D$ has both non-diagonal and diagonal matrix elements from this view point. Besides the $D\bar{D}$ as the first channel in the matrix, we include the second $D^*\bar{D}+c.c$ channel in our calculation.

The vertex $\Gamma^D$ contains the diagonal self-energy $U^D$, which couples the channels to the resonances. The couplings are treated here as free point parameters while in Ref.~\cite{CaoFano2014} we used microscopic quark model interaction form factors. Hence, the matrix element $\Gamma_1^D$ contains \emph{elastic resonances} produced elastically in the $D\bar D$ kinematical s-channel. Thus, we identify the expectation value $\Gamma^D_1$ in the $D\bar D$ channel with the form factor $F_{c}$ introduced in Eq.~(\ref{eq:BWpsi3770}). Rather than invoking a dynamical model for the $D\bar D$ final state interactions, at this stage we parameterize directly the vertex $\Gamma_1^D$ in the form of Eq.~(\ref{eq:BWpsi3770}) and $\Gamma_2^D = 1$ which contains only inelastic resonances, e.g. the $\psi(3686)$ and $\psi(4040)$. The unitarity of Eq.~(\ref{eq:newFDD2}) could be restored by using the Omn\`{e}s function \cite{Omnes:1958} instead of Eq.~(\ref{eq:BWpsi3770}), for example used in Ref.~\cite{Hanhart2012}. However, for the present purpose the Breit-Wigner parametrization of the form factor in Eq.~(\ref{eq:BWpsi3770}) is sufficient because of the narrow width of $\psi(3770)$, much smaller than, e.g. for the $\rho$-meson. We also note that the use of the Omn\`{e}s function would require as input the hitherto unknown $P$-wave $D\bar{D}$ phase shift up to high energies, see e.g. \cite{YRLiu2010}.

The purely diagonal dressed channel propagator $G^D\Gamma^D$ is treated in parameterized form, which follows closely the convention widely used in the literature. Considering that $G^D\Gamma^D =\left((G^{D})^{-1}-U^D \right)^{-1}$ we find
\be
Im\left( G^D\Gamma^D \right)=\frac{Im\left(U^D\right)}{\left( (G^{D})^{-1}-Re\left(U^D\right)\right)^2+Im^2\left(U^D \right)}
\ee
factorized essentially into the imaginary part of the self-energy $U^D$ and a reaction form factor containing all elastic interactions in the channel under consideration. Using the relation between the imaginary part of the self-energy and the decay width the imaginary part of the $D\bar D$ partial wave propagator is found as
\be\label{eq:ImG}
Im\left(G^D\Gamma^D \right)_i= -
\frac{s}{6\pi}\sum_i{\left(\frac{p_{i}(s)}{\sqrt{s}}\right)^3\theta(s-s_i)}|\Gamma^D_i(s)|^2
\ee
with $s_i$ being the threshold energy of the $i$-th channel \cite{fnote:ImG}. The summation runs over the isospin states contributing to the partial widths in channel $i$. The momenta $p_{1}=p_{d}=p_{0,\pm}$ are defined as before in sect.\ref{sec:FanoQuark} and
\bea\label{eq:pcm2}
p_{2}(s) = \sqrt{\frac{\left(s-(m_{D^*}+m_{D})^2\right)\left(s-(m_{D^*}-m_{D})^2\right)}{4s}}.
\eea
As discussed above, the remaining vertex and propagator form factors in channel $i=1$ are taken care of by the (dimensionless) form factor $\Gamma^D_1=F_c$, introduced in Eq.~(\ref{eq:BWpsi3770}). In the $D^*\bar D +c.c.$ channel ($i=2$) we use $\Gamma^{D}_{2} = 1$, i.e. elastic interactions in this channel are neglected. Using subtracted dispersion relations, we obtain the dressed propagators by the following dispersion integrals:
\bea
\left(G^D\Gamma^D \right)_1 &=&\eta_1+ \frac{s}{\pi} \int_{s_1}^\infty \frac{ds'}{s'}\frac{Im\left(G^D\Gamma^D\right)_1}{s'-s-i\epsilon}
\label{eq:Sigma1} \\
\left(G^D\Gamma^D \right)_2 & =& \eta_2 s + \frac{s^{2}}{\pi} \int_{s_2}^\infty{\frac{ds'}{s'^{2}}\frac{Im\left(G^D\Gamma^D\right)_2}{s'-s-i\epsilon}}\label{eq:Sigma2}
\eea
In both cases we have chosen $s=0$ as subtraction point, corresponding to the constraint that the self-energies and correspondingly the propagators should vanish at $s=0$. The twice-subtracted dispersion relation, Eq.~(\ref{eq:Sigma2}), resembling the Roy-equation \cite{Roy:1971}, was found to improve considerably the convergence and stability of the numerical results. The subtraction constant in Eq.(\ref{eq:Sigma1}) is absorbed into the channel mass parameter, corresponding to choose $\eta_1=0$ in the following. The same argument applies to a constant subtraction term in Eq.(\ref{eq:Sigma2}) which has been left out from the beginning. The remaining non-trivial subtraction constant $\eta_2$ is discussed below.

The matrix elements of the induced non-diagonal self-energy is taken to be
\be
W^D_{ij} = \sum_r \frac{g^{r*}_i g^r_j}{s-m^2_{r}} \quad ,
\ee
giving rise to channel coupling through intermediate excitations of $c\bar c$ states. The first ($r=1$) and the second ($r=2$) resonance being the (sub-threshold) $\psi(3686)$ and the $\psi(4040)$ state, respectively. According to Eq.~(\ref{eq:effVertex}) those resonances also contribute to the effective production vertex which is taken into account by
\be \label{eq:Mkpara}
\tilde{M}^D_k(s) = c_{k} + s\sum_r{ \frac{g^{r*}_k g^r_{\psi\gamma}}{s-m^2_{r}}} \quad .
\ee
abandoning now the index $\tau$ with the understanding that $\tau=e^-e^+$ is considered and denoting here and in the following the explicitly observed $D\bar D$ channel by $k=d=1$. We have included the spectroscopic amplitude $x_d$ into the effective production vertex $c_k\equiv x_dM_{d\tau}$ and the coupling constants $g^l_k$ which, as a result, in principle may depend weakly on energy. Using $c_1 =1$ ensures that in Eq.~(\ref{eq:newFDD2}) the bare $\psi(3770)$ form factor $\Gamma^D = F_c$ is retained. The linear dependence on $s$ complies with gauge invariance of the $e^-e^+\to\gamma^*\to \psi$ production vertex.

\begin{figure*}
\begin{center}
{\includegraphics*[width=10.cm]{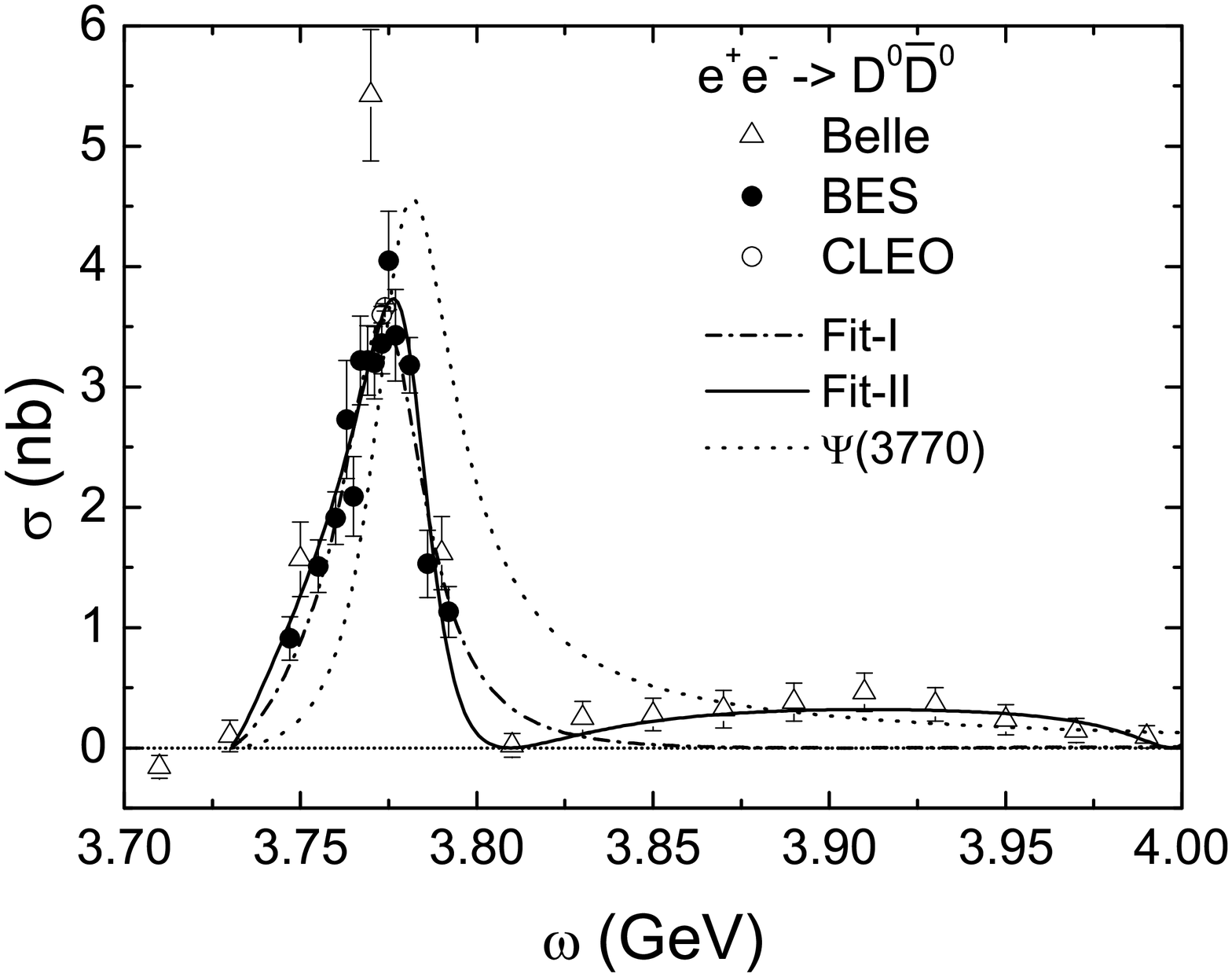}}
{\includegraphics*[width=10.cm]{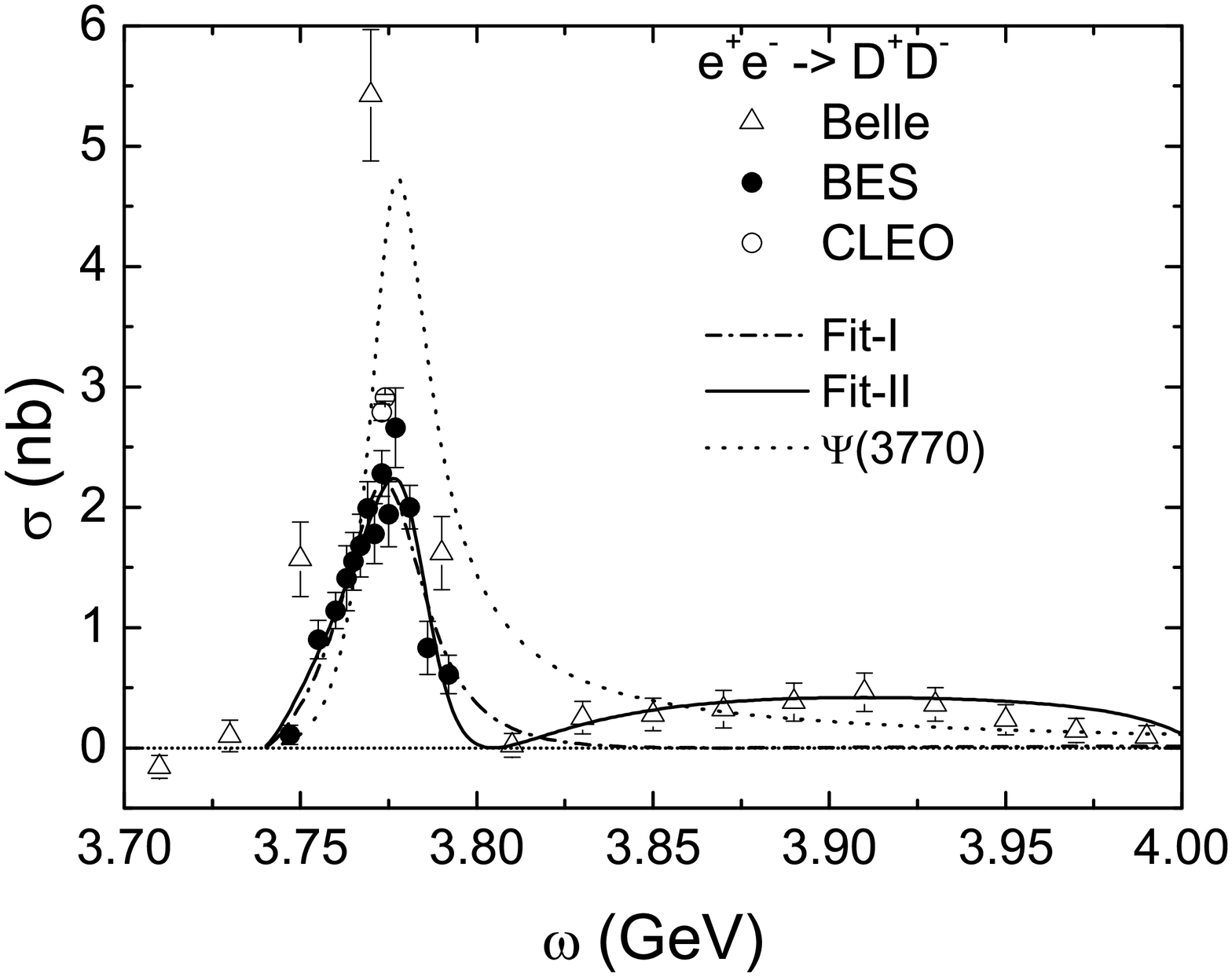}}
\caption{
Total cross section of $e^+e^- \to D^0\bar{D}^0$ (upper) and $e^+e^- \to D^+D^-$ (lower) reactions in parameterized coupled-channel model. The dotted curve is the bare $\psi(3770)$ contribution in Fit-II. The data are from BES~\protect\cite{BESDDbar}, Belle~\cite{BelleDDbar}, and CLEO~\cite{CLEODDbar} collaborations.
\label{tcsdiagrBES}}
\end{center}
\end{figure*}

With the pieces in Eqs.~(\ref{eq:ImG}),~(\ref{eq:Sigma1}),~(\ref{eq:Sigma2}) and ~(\ref{eq:Mkpara}),
the production cross section is obtained from the modulus squared of the reaction amplitude, Eq.~(\ref{eq:newFDD2}), and appropriate phase space factors. In order to indicate the relation to the previous sections we slightly rearrange terms. Using $\Omega^D=1+\Omega^D W^D G^D\Gamma^D$ we extract the $d$-channel $D\bar{D}$ direct production amplitude $F^D_{\tau,d}$ in the first ,
and obtain the partial wave cross section as
\bea \label{eq:tcsTmatrix}
\sigma_d(s) &=& \frac{8\pi \alpha^2 p^3_d}{3s^{5/2}}|F^D_{\tau,d}|^2 \nonumber \\
&=& \frac{8\pi \alpha^2 p^3_d}{3s^{5/2}}|\Gamma^D_{d} \tilde{M}^D_k + \Gamma^D_{d} \sum_k{ \left(\Omega^D_{d,k} -\delta_{kd}\right) \tilde{M}^D_k }|^2 \quad\quad
\eea
Different to the scheme of section \ref{sec:FanoQuark} and section \ref{sec:TheoryCC}, we see that in the usual T-matrix approach the $c\bar c$ configurations are contributing not as separate components but only as intermediate states. The direct $D\bar D$ amplitude, $F_dM_{d\tau}$, contained in the first term interferes with the resonance pieces contained also in $M_d$ and in the re-scattering contributions. This is a quite different explanation for the distortion of the line shapes of the spectral distributions, interpreting asymmetric line shapes solely on the level of leading and next-to-leading order processes through the $D\bar D$ channel space. The results below seem to support the success of this widely used approach, but we must keep in mind the rather biased physical picture behind.

\begin{table*}[t]
\begin{center}
\begin{tabular}{|c|c|c|c|c|}
\hline
& \multicolumn{2}{c|}{Fit-I} & \multicolumn{2}{c|}{Fit-II} \\
\hline
& $D^0\bar{D}^0$ & $D^+D^-$ & $D^0\bar{D}^0$ & $D^+D^-$ \\
\hline
$m_{\psi'}$ (MeV) & 3783.2 $\pm$ 2.5 & 3782.5 $\pm$ 2.9 & 3787.0 $\pm$ 2.2 & 3787.8 $\pm$ 3.5\\
$g_{\psi'D\bar{D}}$ & 14.1 $\pm$ 0.7 & 14.5 $\pm$ 0.5 & 11.6 $\pm$ 0.2 & 11.6 $\pm$ 0.1\\
$g^1_1 = g_{\psi(3686)D\bar{D}}$ & 0.17 $\pm$ 0.02 & 0.17 $\pm$ 0.03 & 0.17 $\pm$ 0.03 & 0.17 $\pm$ 0.05\\
$g^1_2 = g_{\psi(3686) D^*\bar{D}}$ & --- & --- & 1.19 $\pm$ 0.03 & 1.29 $\pm$ 0.06\\
$g^2_1 = g_{\psi(4040) D\bar{D}}$ & --- & --- & 0.145 $\pm$ 0.014 & 0.160 $\pm$ 0.021\\
$g^2_2 = g_{\psi(4040) D^*\bar{D}}$ & --- & --- & 0.38 $\pm$ 0.09 & 0.27 $\pm$ 0.11\\
$g^1_{\psi\gamma} = g_{\psi(3686) \gamma}$ & 6.4 $\pm$ 0.5 & 6.7 $\pm$ 0.9 & 7.6 $\pm$ 0.1 & 8.0 $\pm$ 0.2 \\
$g^2_{\psi\gamma} = g_{\psi(4040) \gamma}$ & --- & --- & 3.4 $\pm$ 0.4 & 2.9 $\pm$ 0.7\\
$c_2$ & --- & --- & -13.9 $\pm$ 1.4 & -17.4 $\pm$ 1.2 \\
$\eta_2$ & --- & --- & 0.39 $\pm$ 0.19 & 0.41 $\pm$ 0.16 \\
$\chi^2/d.o.f$ & 1.12 & 0.92 & 1.43 (1.00$^*$) & 1.57(1.25$^*$) \\
\hline
\end{tabular}
\end{center}
\caption{Fitted parameters in different fit schemes.\\
$^*$: excluding one data point of Belle at $\sqrt{s} = 3.770$ GeV.
\label{Tab:fitpara}}
\end{table*}

Numerical calculations from Eq.~(\ref{eq:tcsTmatrix}) have been performed for the electro-production of charmonium in the mass region of $\psi(3770)$.  We include $D\bar D\in \{D^0\bar D^0,D^- D^+\}$, denoted by $k=d=1$, and $D^*\bar D$, $D\bar D^*$ channels, denoted by $k=2$. The usual Breit-Wigner form is taken for the $D\bar D$ form factor $F_d$. The data~\cite{BESDDbar,BelleDDbar} show two dips in the cross section of $e^+e^- \to D\bar{D}$ below 4.0~GeV. From previous studies it is known that the $P$-wave $D\bar{D}$ interaction is weak~\cite{YRLiu2010}, and also the two relevant channels are weakly coupled. As a result, within the restricted $D\bar D$ T-matrix approach these dips cannot be explained from dynamics. We could speculate, however, that the dips are originating from the electro-production vertices $\tilde{M}_k$. Accepting this point of view, the coupling constants, and therefore the production matrix elements, are largely determined by the location of the interference minima in the cross section.

We have tested two models. First, we explore a reduced problem (Fit-I) modelling the data with $\sqrt{s} \leq $ 3.8~GeV from BES collaboration (14 points)~\cite{BESDDbar} both in the neutral and charged $D\bar{D}$ channels. Including only the $D\bar{D}$ channel and $\psi(3686)$ state and using the once subtracted self-energies, see Eq.~(\ref{eq:Sigma1}), the number of free parameters reduces to four. Results are displayed in Fig.~\ref{tcsdiagrBES} and Tab.~\ref{Tab:fitpara}. As shown, we have achieved an excellent description of the line shape of the $\psi(3770)$, whose $\chi^2/d.o.f$ are considerably better than in previous isobar model analyses~\cite{HBLi2010,Zhang2010}.
Next, we consider the Belle (13 points)~\cite{BelleDDbar} and CLEO (2 points)~\cite{CLEODDbar} data sets for $\sqrt{s} \leq $ 4.0~GeV by calculations additionally including the $\psi(4040)$ state and using the twice subtracted dispersion relations, Eq.~(\ref{eq:Sigma2}) in the $D^*\bar{D} + c.c$ channel (Fit-II). A reasonable agreement with the data is achieved, however, slightly worse than achieved before in Fit-I, as indicated by $\chi^2/d.o.f = 1.43$. It is puzzling that the data point of the Belle collaboration at $\sqrt{s} = 3.770$ GeV is obviously much higher than the nearby data points of BES and CLEO collaborations, see Fig.~\ref{tcsdiagrBES}. If we disregard that data point, the fit is improved with a significantly decreased $\chi^2/d.o.f=1.00$, see Tab.~\ref{Tab:fitpara}. In Fig.~\ref{tcsdiagrBES}, our results are summarized together with the bare line shape of the $\psi(3770)$. Not only the $\psi(3770)$ line shape is properly described, but also the broad enhancement around 3.90 GeV is explained equally well. It is fully determined by the tail of the $\psi(3770)$ bending by the $D^*\bar{D} + c.c$ threshold opening and the onset of the $\psi(4040)$, leaving little room for a real resonance, quoted in the literature as $X(3900)/G(3900)$~\cite{HBLi2010}. This is illuminating for the study of the cross sections of other channels, e.g. the $e^+e^- \to J/\psi\pi^0$ and $J/\psi\eta$ reactions~\cite{Wang:2011yh}. Our extracted parameters in the neutral and charged $D\bar{D}$ channels are consistent with each other within the uncertainties, as can be seen in Tab.~\ref{Tab:fitpara}. This confirms the hypothesis that the distinct behavior of the $D^0\bar{D}^0$ and $D^+D^-$ channels could be largely explained by the different phase space caused by the mass gap between neutral and charged $D$-mesons~\cite{Rosner2005,Voloshin2005}. It should be noted that $m_{\psi'}$ can be viewed as the bare mass of the $\psi(3770)$. Its fitted value is a little larger than the isobar model mass \cite{HBLi2010,Zhang2010}, but consistent with that in the parameterized coupled-channels K-matrix model~\cite{Achasov2012}. The pole of the $\psi(3770)$ is found to be stable in the complex plane at the location (3778.2 $\pm$ 3.8) MeV + i (13.9 $\pm$ 1.4) MeV in both fits, indicating its robust properties.

The mass of $\psi(3686)$ is kept fixed in the above fits. If we allow it to vary freely, we find that it is highly uncertain with a value of 3716.0 $\pm$ 30.0 MeV, whose upper bound is consistent with the value of the above Fano-type formula. Though its lower bound is compatible with the mass of $\psi(3686)$, the existence of a hidden $1^{--}$ charmonium state as indicated by lattice calculation~\cite{LQCD2008} is not excluded by our analysis. Moreover, the comparatively large uncertainty probably indicates self-energy contributions induced by continuum couplings, e.g.~\cite{CaoFano2014}.

\section{Summary} \label{sec:Summary}

In summary, we have developed a formalism for charmonium and other production reactions. The theoretical formulation has been kept general and is providing a scheme for the consistent description of open and closed channels, accounting also for sub-threshold states. As a particular new aspect, we have considered in detail the case of channels closed by quark confinement. Such channels are of natural importance for quarkonium production. For that goal we have taken a slightly different view, namely considering the well known open-closed channel concept not from the hadron- but from the quark-side. The connection to the Fano-approach was worked out which in other fields of physics is a well established and successful approach and gives us a simple and direct picture of the interference phenomenon. We have analysed the $e^+e^- \to D^0\bar{D}^0$ and $e^+e^- \to D^+D^-$ reactions and extracted the dynamical line shape parameters for the $D^0\bar D^0$ and the $D^-D^+$ channels, leading to the conclusion that the $\psi(3770)$ state is a resonance embedded in the $D\bar{D}$ continuum. For comparison we have repeated a conventional T-matrix analysis, including coupled-channels effects up to the energy range above the $\psi(3770)$ state. Besides a good description of the $\psi(3770)$ spectral distribution we find in our parameterized coupled-channel formalism that the broad $X(3900)/G(3900)$ structure is of non-resonant nature. In the coupled channels scheme, discussed in section \ref{sec:TheoryCC} and the closely related Fano-picture, it is understood in a natural way that the line shape of the $\psi(3770)$ state varies from one to the other production and decay channel because of the different non-resonant continua and interfering contributions. Hence, in view of the above results we may draw an almost trivial and rather obvious conclusion, namely that line shapes are fingerprints of the special dynamical conditions under which a state was produced and is decaying. This is intriguing for our understanding of the properties of many other hadronic states, especially with respect to the newly found and hardly understood X,Y,Z states.

\begin{acknowledgments}

This work was supported by the Deutsche Forschungsgemeinschaft (CRC16, Grants No. B7 and No. Le439/7) and in part by I3HP SPHERE, the LOEWE and the National Natural Science Foundation of China (Grant Nos. 11347146 and 11405222).

\end{acknowledgments}


\begin{thebibliography}{1}
%
\bibitem{BES:2013}M. Ablikim {\it et al}. (BES Collaboration), Phys. Rev. Lett. \textbf{110}, 252001 (2013).
%
\bibitem{Belle:2013}Z. Liu {\it et al}. (Belle Collaboration), Phys. Rev. Lett. \textbf{110}, 252002 (2013).
%
\bibitem{CLEOc:2013}T. Xiao, S. Dobbs, A. Tomaradze and K. K. Seth, Phys. Lett. \textbf{B 727}, 366 (2013).
%
\bibitem{BES:1896}M. Ablikim {\it et al}. (BES Collaboration), Phys. Rev. Lett. \textbf{111}, 242001 (2013).
%
\bibitem{BES:2760}M. Ablikim {\it et al}. (BES Collaboration), Phys. Rev. Lett. \textbf{112}, 132001 (2014).
%
\bibitem{Otto2013}C. Ott {\it et al}., Science \textbf{340}, 716 (2013).
%
\bibitem{caoKSigma}Xu Cao, V. Shklyar, and H. Lenske, Phys. Rev. C \textbf{88}, 055204 (2013).
\bibitem{JuliaDiaz:2006is}
  B.~Julia-Diaz, B.~Saghai, T.-S.~H.~Lee and F.~Tabakin,
  Phys.\ Rev.\ C {\bf 73}, 055204 (2006).

\bibitem{Matsuyama:2006rp}
  A.~Matsuyama, T.~Sato and T.-S.~H.~Lee,
  Phys.\ Rept.\  {\bf 439}, 193 (2007).
%
\bibitem{FanoPR1961}U. Fano, Phys. Rev. \textbf{124}, 1866 (1961).
%
\bibitem{Orrigo:2005}S. E. A. Orrigo, H. Lenske, and F. Cappuzzello {\it et al}., Phys. Lett. \textbf{B633}, 469 (2006).
%
\bibitem{CaoFano2014}Xu Cao and H. Lenske, arXiv:1408.5600 [nucl-th].
%
\bibitem{Eichten1980}E. Eichten, K. Gottfried, T. Kinoshita, K. D. Lane and T. M. Yan, Phys. Rev. D \textbf{17}, 3090 (1978); \textbf{21}, 203 (1980).
%
\bibitem{klshnkv2005}Yu. S. Kalashnikova, Phys. Rev. D \textbf{72}, 034010 (2005).
%
\bibitem{Barnes2008}T. Barnes, S. Godfrey and E. S. Swanson, Phys. Rev. D \textbf{72}, 054026 (2005).
%
\bibitem{BESDDbar}M. Ablikim {\it et al}. (BES Collaboration), Phys. Lett. \textbf{B668}, 263 (2008).
%
\bibitem{BelleDDbar}G. Pakhlova {\it et al}. (Belle Collaboration), Phys. Rev. D \textbf{77}, 011103(R) (2008).
%
\bibitem{CLEODDbar}Q. He {\it et al}. (CLEO Collaboration), Phys. Rev. Lett. \textbf{95}, 121801 (2005); S. Dobbs {\it et al}. (CLEO Collaboration), Phys. Rev. D \textbf{76}, 112001 (2007).
%
\bibitem{BEShadrons}M. Ablikim {\it et al}. (BES Collaboration), Phys. Rev. Lett. \textbf{101}, 102004 (2008).
%
\bibitem{KEDRhadrons}V. V. Anashin {\it et al}., Phys. Lett. \textbf{B711}, 292 (2012).
%
\bibitem{HBLi2010}H. B. Li, X. S. Qin, and M. Z. Yang, Phys. Rev. D \textbf{81}(R), 011501 (2010).
%
\bibitem{Zhang2010}Y. J. Zhang and Q. Zhao, Phys. Rev. D \textbf{81}, 034011 (2010).
%
\bibitem{YRLiu2010}Y. R. Liu {\it et al}., Phys. Rev. D \textbf{82}, 014011 (2010).
%
\bibitem{Achasov2012}N. N. Achasov and G. N. Shestakov, Phys. Rev. D \textbf{86}, 114013 (2012); Phys. Rev. D \textbf{87}, 057502 (2013).
%
\bibitem{Chen2013}G. Y. Chen and Q. Zhao, Phys. Lett. \textbf{B718}, 1369 (2013).
%
\bibitem{Novikov:1978}V. A. Novikov \emph{et al.}, Phys. Rept. \textbf{41},1 (1978).
%
\bibitem{pdg2010}J. Beringer {\it et al}. (Particle Data Group), Phys. Rev. D \textbf{86}, 010001 (2012).
\bibitem{Aceti:2014ala}
  F.~Aceti, L.~R.~Dai, L.~S.~Geng, E.~Oset and Y.~Zhang,
  Eur.\ Phys.\ J.\ A {\bf 50}, 57 (2014).

\bibitem{Hyodo:2011qc}
  T.~Hyodo, D.~Jido and A.~Hosaka,
  Phys.\ Rev.\ C {\bf 85}, 015201 (2012).

\bibitem{Hyodo:2007jq}
  T.~Hyodo and W.~Weise,
  Phys.\ Rev.\ C {\bf 77}, 035204 (2008).

\bibitem{Hyodo:2013nka}
  T.~Hyodo,
  Int.\ J.\ Mod.\ Phys.\ A {\bf 28}, 1330045 (2013).
%
\bibitem{Hanhart2012}C. Hanhart, Phys. Lett. \textbf{B715}, 170 (2012).
%
\bibitem{zhangPRL09}Y. J. Zhang, G. Li and Q. Zhao, Phys. Rev. Lett. \textbf{102}, 172001 (2009).
%
\bibitem{Wang:2011yh}Q.~Wang, X.~H.~Liu and Q.~Zhao, Phys.\ Rev.\ D {\bf 84}, 014007 (2011).
%
\bibitem{Omnes:1958}R. Omn\`{e}s, Nuov. Cim. \textbf{8}, 316 (1958).
%
\bibitem{fnote:ImG}{Eq.(\ref{eq:ImG}) corresponds, in fact, to a self-consistency problem which is treated here in lowest order only.}
%
\bibitem{Roy:1971}S. M. Roy, Phys. Lett. 36B, (1971) 353.
%
\bibitem{Rosner2005}J. L. Rosner, Ann. Phys. \textbf{319}, 1 (2005).
%
\bibitem{Voloshin2005}M. B. Voloshin, Phys. Atom. Nuc. \textbf{68}, 771 (2005).
\bibitem{LQCD2008}J. J. Dudek and E. Rrapaj, Phys. Rev. D \textbf{78}, 094504 (2008); L. Liu {\it et al}., JHEP \textbf{07}, 126 (2012).

\end{thebibliography}
\end{document}